\documentclass[aps,prd,reprint,amsmath,amssymb,superscriptaddress,longbibliography,nofootinbib]{revtex4-2}
\usepackage[T1]{fontenc}
\usepackage{lmodern}
\usepackage{bm}
\usepackage{mathrsfs}
\newcommand{\unreferencedlabel}[1]{}
\usepackage{xcolor}
\usepackage[colorlinks=true,citecolor=blue,linkcolor=blue,urlcolor=blue]{hyperref}
\hypersetup{pdftitle={First law of thermodynamics for interacting charged black holes and applications},pdfauthor={Qing Han, Marcus Khuri, Gilbert Weinstein, Jingang Xiong}}
\newtheorem{theorem}{Theorem}
\newtheorem{proposition}[theorem]{Proposition}
\newcommand{\dd}{\mathrm d}
\newcommand{\R}{\mathbb R}
\newcommand{\cE}{\mathscr E}
\newcommand{\cA}{\mathcal A}
\newcommand{\HH}{\mathsf H}
\newcommand{\II}{\mathsf I}
\newcommand{\Eflat}{\widehat{\boldsymbol E}}
\begin{document}
\title{First law of thermodynamics for interacting charged black holes and applications}
\author{Qing Han}
\affiliation{Department of Mathematics, University of Notre Dame, Notre Dame, Indiana 46556, USA}
\author{Marcus Khuri}
\affiliation{Department of Mathematics, Stony Brook University, Stony Brook, New York 11794, USA}
\author{Gilbert Weinstein}
\affiliation{Department of Mathematics and Department of Physics, Ariel University, Ariel 40700, Israel}
\author{Jingang Xiong}
\affiliation{School of Mathematical Sciences, Laboratory of Mathematics and Complex Systems, MOE, Beijing Normal University, Beijing 100875, China}
\begin{abstract}
We derive a mass variation identity expressing the first law of black hole thermodynamics for asymptotically flat, stationary, axisymmetric Einstein–Maxwell solutions with multiple electrically charged horizons and finite conical axis singularities. Under the stated asymptotic and parameter regularity hypotheses, the identity holds for independent variations of all parameters within families of fixed horizon types. On the all-degenerate family, a correction to the ADM mass defines a scalar function whose derivatives with respect to angular momenta are the horizon angular velocities. At the Majumdar–Papapetrou family, its angular momentum Hessian is positive definite. Combining the first variation identity with a separate theorem on the sign of the conical parameters shows that this Hessian is nondecreasing in the positive-semidefinite order along admissible paths of increasing Weyl gaps at fixed positive charges.
\end{abstract}
\maketitle

\section{Introduction}

The first law of black hole thermodynamics is an equality of one-forms on a parameter space of stationary solutions. For one horizon, the differential of the ADM mass matches changes in horizon area, angular momentum, and electric charge with the surface gravity, angular velocity, and electric potential respectively~\cite{BCH,Wald,IyerWald,Gao}. For several horizons separated by conical axis singularities, the differential also contains terms depending on the cone angles. The problem is to compute these coefficients for variations of the prescribed geometric and potential data.

First laws with conical terms are known for charged static binaries \cite{KZ}, rotating vacuum binaries \cite{RGM}, rotating charged binaries \cite{GMR}, and static arrays \cite{GLS}. Cl\'ement and Gal'tsov formulated a general multi-horizon law, including magnetic and gravitomagnetic sectors \cite{CG}. Note however, that in Sec.~3 of Ref.~\cite{CG}, the first law is motivated by comparing Euler's homogeneity identity with the Smarr relation. This comparison tests the first law only along uniform rescaling and does not by itself determine its coefficients in all parameter directions. In particular, uniform scaling leaves the cone angles unchanged and therefore does not determine the coefficient of the conical term in the first law. However, they obtain the conical term through a NUTless limiting calculation in Sec.~4 of Ref.~\cite{CG}. Here we give an independent derivation for electrically charged black holes separated by uncharged conical struts, establishing the first law in all admissible parameter directions. We achieve this by analyzing the boundary terms after integrating by parts appropriately renormalized energies. We then use it to compare second derivatives of a corrected mass function.

We carry out this calculation under the analytic hypotheses in Sec.~\ref{sec:input}. Horizon ordering and types are fixed, while locations, rod lengths, and independent axis potential constants are allowed to vary. On the all-degenerate family, integration by parts of the derivative of a renormalized harmonic map energy gives the derivatives with respect to angular momenta and charges. A stress tensor boundary integral gives the derivatives with respect to puncture positions. For nondegenerate horizons, a further static reference reduction of the energy yields the first law in full generality.

For the application, we introduce the scalar function
\begin{equation}
 \cE=m-\frac14\sum_{j=2}^N L_j\alpha_j,
 \label{eq:potentialdefinition}
\end{equation}
which we call the corrected mass function. Here $L_j$ is the Weyl coordinate length of the $j$-th finite axis rod and $\alpha_j$ is the constant value of the Weyl metric coefficient $\alpha$ on that rod, which also measures its logarithmic angle defect. On the all degenerate family, parametrized by $(L,J,Q)$, the first law gives
\begin{equation}
 \dd\cE=\sum_i(\Omega_i\dd J_i+\Phi_i\dd Q_i)
                    -\frac14\sum_j\alpha_j\dd L_j.
 \unreferencedlabel{eq:introgradient}
 \notag
\end{equation}
Thus the horizon angular velocities $\Omega_i$ are the components of $\nabla_J\cE$, where $J_i$ are the angular momenta.

Fix positive charges and consider the Majumdar--Papapetrou family, where $J=0$ and $\cE=\sum_iQ_i$ independently of $L$. Define its spin Hessian by $\HH(L,Q)=D_J^2\cE(L,0,Q)$. Equivalently, $\HH$ is the Jacobian of $J\mapsto\Omega$ at fixed charges and gaps. Subject to the hypotheses of Sec.~\ref{sec:response}, we obtain
\begin{equation}
 \HH(L,Q)>0,\qquad \HH(L',Q)-\HH(L,Q)\geq0
 \unreferencedlabel{eq:introresponse}
 \notag
\end{equation}
when $L'_j\geq L_j$ for every $j$. For symmetric matrices, $A>0$ means positive definite and $A\geq0$ means positive semidefinite. The physical interpretation of $\HH^{-1}$ as a linear moment of inertia matrix follows from $\Omega=\HH J+o(|J|)$.

The separation variables must be distinguished from proper distances. The coefficient paired with a cone-angle variation is a redshifted length, defined in Eq.~\eqref{eq:thermolength}. It can be finite even when an extremal throat makes the proper distance infinite. Interaction-dependent angular velocities are familiar from explicit double-Kerr families \cite{CHR}; the present comparison concerns an arbitrary number of charged horizons and fixed individual charges.

The paper is organized as follows. Section~\ref{sec:setting} specifies the solution family and the asymptotic and variation hypotheses, and states the first law as Theorem~1. Section~\ref{sec:energy} treats degenerate horizons by varying the renormalized energy with respect to the axis potential data and puncture positions. Section~\ref{sec:covariant} adds nondegenerate horizons. We further reduce the renormalized energy by subtracting a static reference solution, then adapt the mass identity of Khuri, Sokolowsky, and Weinstein to several horizon rods and conical axes. We then evaluate the moving pole contributions to the energy variation and combine them with the horizon and static reference variations to prove the first law. Section~\ref{sec:potential} studies the corrected mass function and its derivatives. Section~\ref{sec:response} proves positivity and comparison results for its angular momentum Hessian. Appendix~\ref{app:boundary} explains the boundary limits used in the variation, Appendix~\ref{app:horizon} computes the gravitational contribution at a horizon, and Appendix~\ref{sec:binarycheck} checks the first law explicitly for the Bach--Weyl two body solution.

\section{Geometric setting and hypotheses}
\label{sec:setting}

\subsection{Horizons, charge, and conical variables}

We work in four spacetime dimensions with $G=c=1$, signature $(-+++)$, and Einstein--Maxwell Lagrangian
\begin{equation}
 \boldsymbol L=\frac{R}{16\pi}\boldsymbol\epsilon
                 -\frac1{8\pi}F\wedge\star F,
 \qquad F=\dd\cA.
 \unreferencedlabel{eq:lagrangian}
 \notag
\end{equation}
Here $R$ is the scalar curvature of the spacetime metric, and $\star$ is its Hodge star operator.
Let $\tau$ be the stationary Killing field normalized at infinity, and let $\eta$ generate rotations with period $2\pi$. The exterior is asymptotically flat and has $N$ horizon components, with $N-1$ finite axis rods between them. The open orbit half-plane is simply connected, with the usual rod topology and no additional cycles obstructing the axial potentials. The two unbounded axis rods are regular. All finite axis defects are purely conical: there are no time-shift defects, additional exterior singularities, or charged material lines.

We assume that $F=\dd\cA$ for a globally defined one-form $\cA$, invariant under $\tau$ and $\eta$ and regular on the open axes and in coordinates extending across the future horizons. We normalize $\cA(\tau)\to0$ at infinity and require the potential and its first derivatives with respect to the solution parameters to be $O(r^{-1})$ in asymptotically Cartesian components. In particular, every horizon has zero magnetic charge, although the magnetic field need not vanish. Solutions with nonzero magnetic charges require local potentials and additional boundary terms arising from their gauge transitions \cite{CG,Prabhu}.

The generator of horizon $i$ is $\xi_i=\tau+\Omega_i\eta$. Its area, surface gravity, charge, and electric potential are
\begin{equation}
 A_i,\qquad \kappa_i,\qquad
 Q_i=\frac1{4\pi}\int_{H_i}\star F,\qquad
 \Phi_i=-\xi_i\cdot\cA\big|_{H_i}.
 \unreferencedlabel{eq:horizondata}
 \notag
\end{equation}
Here $H_i$ denotes a horizon cross-section in surface integrals. The surface gravity is defined by $\nabla_{\xi_i}\xi_i=\kappa_i\xi_i$ on the horizon. The contraction in $\Phi_i$ evaluates the one-form $\cA$ on $\xi_i$: in adapted coordinates it is $\cA_t+\Omega_i\cA_\phi$. The quantities $\Omega_i$, $\kappa_i$, and $\Phi_i$ are constant on each connected component of the horizon. The appropriate horizon angular momentum includes the electromagnetic contribution:
\begin{equation}
 J_i=\frac1{16\pi}\int_{H_i}\star\dd\eta^\flat
       +\frac1{4\pi}\int_{H_i}(\eta\cdot\cA)\star F.
 \label{eq:angularmomentum}
\end{equation}

In adapted coordinates $\tau=\partial_t$ and $\eta=\partial_\phi$; $t$ denotes coordinate time. Write the Weyl--Papapetrou metric as
\begin{equation}
 \dd s^2=-e^{2U}\dd t^2+e^{-2U}\!\left[
 \rho^2(\dd\phi+w\dd t)^2+e^{2\alpha}(\dd\rho^2+\dd z^2)\right]\!.
 \unreferencedlabel{eq:metric}
 \notag
\end{equation}
The determinant of the Killing-field Gram matrix fixes $\rho$; the asymptotic normalization fixes the scale of the Weyl coordinates $(\rho,z)$. Ordered horizon rods have endpoints $z_i^-\leq z_i^+$, with $z_i^-=z_i^+=z_i$ for degenerate horizons. The finite gaps are
\begin{equation}
 \Gamma_j=(z_{j-1}^+,z_j^-),\qquad
 L_j=z_j^--z_{j-1}^+>0,\qquad 2\leq j\leq N.
 \unreferencedlabel{eq:gaps}
 \notag
\end{equation}
On each axis rod $\alpha$ has a constant value $\alpha_j$. The outer values are $\alpha_1=\alpha_{N+1}=0$. Define
\begin{equation}
 C_j=e^{-\alpha_j},\quad
 \mu_j=\frac{C_j-1}{4},\quad
 \lambda_j=L_je^{\alpha_j}=\frac{L_j}{1+4\mu_j}.
 \unreferencedlabel{eq:conicalvariables}
 \notag
\end{equation}
The circumference-to-radius ratio is $2\pi C_j$, so the signed deficit angle is $-8\pi\mu_j$. Positive $\mu_j$ is a compressive strut with an angle excess. The tensile string tension is $-\mu_j$.

The thermodynamic length $\lambda_j$ is
\begin{equation}
 \lambda_j=\int_{\Gamma_j}\sqrt{-g_{tt}g_{zz}}\,\dd z
           =\int_{\Gamma_j}e^{\alpha_j}\,\dd z,
 \label{eq:thermolength}
\end{equation}
where the metric is restricted to the axis worldsheet. It is the redshifted proper length, with redshift measured relative to infinity. Our thermodynamic length and compressive force have the opposite signs to the thermodynamic length and tension used in Ref.~\cite{GLS}.

\subsection{Existence and estimates required for the first law}
\label{sec:input}

Existence of harmonic maps with prescribed rod endpoints and axis potential constants has been established in vacuum and electrovacuum for configurations with all horizons nondegenerate and for configurations with all horizons degenerate \cite{Weinstein,WeinsteinEM,KW,HKWXev}. The existence constructions extend to configurations containing both degenerate and nondegenerate horizons with essentially no change.

Let $\mathcal P$ denote the admissible parameter domain of ordered rod endpoints and independent axis potential constants, with horizon types fixed and with the origin of the Weyl $z$-coordinate and the reference values of the axis potentials also fixed. A degenerate horizon is represented by a puncture and satisfies $J_i^2+Q_i^4>0$.

We first specify the two subtractions used in the energy. The harmonic map has components $(u,v,\chi,\psi)$, where $v$ is the twist potential and $\chi,\psi$ are electromagnetic potentials, with the normalization of Sec.~\ref{sec:fields}. The logarithmic renormalization replaces $u$ by $U=u+\log\rho$. For the further reduction, let $(U_*,0)$ be the renormalized vacuum harmonic map of the Bach--Weyl solution with the same nondegenerate horizon rods. If $\mathcal N$ indexes those rods, put $\ell_i=z_i^+-z_i^-$ and
\begin{align}
 r_i^\pm&=\sqrt{\rho^2+(z-z_i^\pm)^2},\notag\\
 U_*&=\frac12\sum_{i\in\mathcal N}
 \log\frac{r_i^++r_i^--\ell_i}{r_i^++r_i^-+\ell_i}.
 \label{eq:staticreference}
\end{align}
The vacuum quadratures determine $\alpha_*$ with $\alpha_*\to0$ at infinity, and $m_*=\frac12\sum_{i\in\mathcal N}\ell_i$. When $\mathcal N=\varnothing$, take $U_*=\alpha_*=m_*=0$. Write $\overline U=U-U_*$ and $\overline\alpha=\alpha-\alpha_*$.
For the harmonic-map calculations, spatial derivatives and norms are
Euclidean. On axisymmetric functions we use
\begin{gather*}
 \Delta_2=\partial_{\rho\rho}+\partial_{zz},\qquad
 \Delta=\Delta_3=\Delta_2+\rho^{-1}\partial_\rho,\\
 \dd^3x=\rho\,\dd\rho\,\dd z\,\dd\phi.
\end{gather*}
With
\begin{align}
 \Theta&=\dd v-\psi\dd\chi+\chi\dd\psi,\notag\\
 P&=e^{4u}|\Theta|^2+e^{2u}(|\nabla\chi|^2+|\nabla\psi|^2),
 \label{eq:potentialdensity}
\end{align}
the renormalized and reduced energies are, respectively,
\begin{align}
 \mathcal E&=\int_{\R^3}(|\nabla U|^2+P)\,\dd^3x,
 \label{eq:energy}\\
 \widehat{\mathcal E}&=\int_{\R^3}(|\nabla\overline U|^2+P)\,\dd^3x.
 \label{eq:multirodenergy}
\end{align}
The first is finite for the all degenerate family; nondegenerate rods require the second subtraction.

\paragraph{Estimates on the solutions and their variations.}

We now list the estimates required to carry out the calculations leading to the first law.
All spatial derivatives and distances below are Euclidean. The $C^0$ and $C^1$ bounds in (B1) refer to $(\rho,z)$ dependence. Dependence on parameters is treated in (B2). Constants are uniform on compact subsets of $\mathcal P$.

\begin{enumerate}
\item[(B1)] \emph{Nonlinear estimates.} At infinity,
\begin{equation}
 \overline U=-\frac{m-m_*}{r}+O_1(r^{-2}),\qquad
 \overline\alpha=O_1(r^{-2}).
 \label{eq:inputinfinity}
\end{equation}
Here $f=O_1(r^{-p})$ if $r^p|f|+r^{p+1}|\nabla f|\leq C$, where $\nabla$ is the spatial gradient. Near a nondegenerate pole, for $F_1=\overline U$, $F_2=\overline\alpha$ and $r_p=|x-pe_z|$, the following holds:
\begin{equation}
 F_j=a_j+H_j,\quad |H_j|+r_p|\nabla H_j|\leq Cr_p^\gamma, \quad j=1,2,
 \label{eq:poleexpansion}
\end{equation}
where $0<\gamma\leq1$, and with the same constant $a_j$ when approaching from the adjoining rods. In addition $P\leq C/r_p$.
At a degenerate puncture,
\begin{equation}
 |\overline U-\log r_p|+|\overline\alpha|\leq C,\qquad
 |\nabla\overline U|+|\nabla\overline\alpha|\leq C/r_p.
 \label{eq:masspuncturebounds}
\end{equation}
These estimates hold uniformly in angle. On compact subintervals of axis and horizon rods, the fields have their prescribed traces and $\rho\overline U_\rho,\rho\overline\alpha_\rho\to0$; on axis rods, $v=c_j$, $\chi=d_j$, and $\psi=O(\rho^2)$.

\item[(B2)] \emph{Linear estimates.}
The fields and their spatial gradients are continuously differentiable
with respect to the parameters away from the axis. Let $x\in\mathbb{R}^3$
denote the position vector in the auxiliary Euclidean space, and let
$e_z$ be the unit vector along the symmetry axis. Along a parameter
curve $s\mapsto X_s$, write $\delta=\partial_s$ for differentiation
at fixed $x$.

Near a moving endpoint $p(s)e_z$, introduce centered coordinates
$y=x-p(s)e_z$ and set $r_p=|y|$. Differentiation at fixed $y$
is given by
\[
 D_s=\partial_s+\dot p\,\partial_z.
\]
At each nondegenerate pole, the coefficients $a_j(s)$ in the
expansions $F_j=a_j+H_j$ of (B1) are continuously differentiable, and
\begin{equation}
 |D_s H_j|+r_p|\nabla D_s H_j|\leq Cr_p^\gamma,
 \qquad j=1,2,
 \label{eq:poleparameterbound}
\end{equation}
\begin{equation}
 |D_s P|\leq C/r_p,
 \label{eq:chargedpolebounds}
\end{equation}
where $0<\gamma\leq1$.
At each degenerate puncture, $D_s\overline U$ is bounded.

On each compact subinterval of an axis rod $\Gamma_j$, as
$\rho\to0$,
\[
 \delta v\longrightarrow\delta c_j,\qquad
 \delta\chi\longrightarrow\delta d_j,\qquad
 \delta\psi=O(\rho^2).
\]

At infinity, the expansions in Eq.~\eqref{eq:inputinfinity}
may be differentiated with respect to the parameters:
\[
 \delta\overline U=-\frac{\delta(m-m_*)}{r}+O_1(r^{-2}),
 \qquad
 \delta\overline\alpha=O_1(r^{-2}).
\]

We also assume the boundary-current estimates and convergence
conditions stated in Appendix~\ref{app:linearconditions}.
These conditions justify the boundary limits and differentiation
of the limiting energies in Sections~\ref{sec:energy}
and~\ref{sec:directreferencevariation}.
\end{enumerate}

\paragraph{Established results and extensions.}
Finiteness of the degenerate renormalized energy is part of the existence theory \cite{KW}. In vacuum, the puncture asymptotics are established in Ref.~\cite{HKWXasym}, and smooth dependence on puncture positions, including the renormalized fields with axis constants fixed, in Ref.~\cite[Theorem~3.1]{HKWXmass}. Li--Tian \cite[Sec.~2, Theorem~2.2]{LiTian93} prove spatial $C^{1,\beta}$ regularity at vacuum poles for every $0<\beta<1$. Nguyen \cite[Theorem~A]{NguyenCMP} proves electrovacuum regularity on the interiors of axis rods. Khuri--Sokolowsky--Weinstein \cite[Lemma~2.1, Sec.~3, and Appendix~A]{KSWPenrose} give pole bounds under smooth minimal boundary and regular axis assumptions.

The vacuum analysis was tailored to degenerate horizons and puncture motion, the most difficult parameter variation. Independent angular momentum variations and motion of nondegenerate poles are expected to require simpler adaptations.

The electrovacuum extensions use the complex hyperbolic plane in place of the hyperbolic plane; its strictly negative sectional curvature suggests the same approach. The charged asymptotics, including those at poles adjoining conical struts, and the additional parameter estimates remain to be verified in forthcoming work \cite{HKWXcomplexprep}.

\subsection{First law}

\begin{theorem}[Mass differential] Let a family over $\mathcal P$ satisfy the geometric conditions of Sec.~\ref{sec:setting} and hypotheses (B1)--(B2). Then, as an equality of one-forms on $\mathcal P$, the ADM mass satisfies
\begin{align}
 \delta m={}&\sum_{i\in\mathcal N}\frac{\kappa_i}{8\pi}\delta A_i\notag\\
 &+\sum_{i=1}^N(\Omega_i\delta J_i+\Phi_i\delta Q_i)
 -\sum_{j=2}^N\lambda_j\delta\mu_j.
 \label{eq:firstlaw}
\end{align}
\end{theorem}
Equivalently, the identity holds after evaluation on the tangent to every differentiable curve in $\mathcal P$. For $i\in\mathcal N$, with $\hbar=k_B=1$, the area term is $\kappa_i\delta A_i/(8\pi)=T_i\delta S_i$, where $T_i=\kappa_i/(2\pi)$ is the Hawking temperature and $S_i=A_i/4$ is the Bekenstein--Hawking entropy of the $i$-th horizon \cite{Hawking1975}. Eq.~\eqref{eq:firstlaw} is an identity pulled back to the solution family. Collisions, disappearing punctures, and transitions between degenerate and nondegenerate horizons lie outside the fixed-type statement.

Section~\ref{sec:energy} proves the first law when all horizons are degenerate. Section~\ref{sec:harmonicnondeg} extends the proof to allow nondegenerate horizons, including configurations with both horizon types.

\section{Variation of the renormalized energy for degenerate horizons}
\label{sec:energy}

This section gives the renormalized energy derivation for degenerate horizons only. We first vary the axis potential data at fixed punctures, then allow the punctures to move, and finally use the relation between renormalized energy and ADM mass to obtain the conical term. The analytic hypotheses used in the boundary limits are collected in Sec.~\ref{sec:input}.

\subsection{Complex-hyperbolic fields and conserved currents}
\label{sec:fields}

For the potentials introduced in Sec.~\ref{sec:input}, we use the normalization
\begin{equation}
 \psi=\eta\cdot\cA,\qquad \dd\chi=\eta\cdot\star F,
 \qquad \Theta=\dd v-\psi\dd\chi+\chi\dd\psi.
 \unreferencedlabel{eq:potentials}
 \notag
\end{equation}
In these conventions $\dd\psi=-\eta\cdot F$ and $\Theta=\tfrac12\star(\eta^\flat\wedge\dd\eta^\flat)$. The orientation is chosen so that the north-minus-south jumps below give positive electric charge and angular momentum. The target metric is
\begin{equation}
 \dd s^2_{\mathbb H^2_{\mathbb C}}
  =\dd u^2+e^{4u}\Theta^2+e^{2u}(\dd\chi^2+\dd\psi^2).
 \unreferencedlabel{eq:target}
 \notag
\end{equation}
This is the Einstein--Maxwell extension of the vacuum harmonic map reduction \cite{Weinstein,WeinsteinEM,KW,HKWXev}.

Regularity of the global Maxwell potential gives $\psi=0$ on every axis rod. If
\begin{equation}
 \begin{gathered}
 v|_{\Gamma_j}=c_j,\qquad \chi|_{\Gamma_j}=d_j,\\
 c_{i+1}-c_i=4J_i,\qquad d_{i+1}-d_i=2Q_i.
 \end{gathered}
 \label{eq:axisjumps}
\end{equation}
then the angular momentum is precisely Eq.~\eqref{eq:angularmomentum}. The residual transformations $\chi\mapsto\chi+a$, $v\mapsto v-a\psi+b$ preserve $\Theta$, the zero axis trace of $\psi$, and the physical jumps. Their telescoping fluxes are unchanged.

The renormalized energy is given by Eq.~\eqref{eq:energy}. The field equations are
\begin{align}
 \Delta U&=2e^{4u}|\Theta|^2
                   +e^{2u}(|\nabla\chi|^2+|\nabla\psi|^2),\unreferencedlabel{eq:Uequation}\notag\\
 \operatorname{div}(e^{4u}\Theta)&=0,\notag\\
 \operatorname{div}(e^{2u}\nabla\chi)&=2e^{4u}\Theta\cdot\nabla\psi,\notag\\
 \operatorname{div}(e^{2u}\nabla\psi)&=-2e^{4u}\Theta\cdot\nabla\chi.
 \unreferencedlabel{eq:potenzialequations}\notag
\end{align}
Writing $A_t=\tau\cdot\cA$, the reconstruction identities needed on the axis are
\begin{equation}
 w_z=-2\rho e^{4u}\Theta_\rho,\qquad
 (A_t)_z=w\psi_z-\rho e^{2u}\chi_\rho.
 \unreferencedlabel{eq:reconstruction}
 \notag
\end{equation}
At an extremal puncture, $w\to-\Omega_i$, and the adjacent-axis limit of $A_t$ is $-\Phi_i$.

\subsection{Spin and charge variations at fixed punctures}

Let $X_s=(U_s,v_s,\chi_s,\psi_s)$ be a differentiable family of harmonic maps with fixed puncture locations, and write $\delta X=\left.\partial_sX_s\right|_{s=0}$. Let $G=G_{\varepsilon,\delta,R}$ be the region inside the ball of radius $R$, with balls of radius $\varepsilon$ around the punctures and the cylinder $\rho\leq\delta$ removed. Write $\mathcal E_G$ for the renormalized energy integrated over $G$. Differentiate this integral with respect to $s$ and integrate by parts. The interior terms are the Euler--Lagrange expressions paired with $\delta X$; they vanish because $X_0$ satisfies the harmonic map equations. Thus only the following boundary integral remains:
\begin{align}
 \delta \mathcal{E}_G=2\int_{\partial G}\big[&\delta U\,\partial_n U
 +e^{4u}\Theta_n(\delta v-\psi\delta\chi+\chi\delta\psi)\notag\\
 &+e^{2u}(\delta\chi\,\partial_n\chi
                         +\delta\psi\,\partial_n\psi)\big]\dd S.
 \label{eq:energyboundary}
\end{align}
Here $n$ is the outward unit normal to $G$; on a puncture sphere it points toward the puncture. Let $\delta\to0$, then $\varepsilon\to0$, and finally $R\to\infty$, as in Appendix~\ref{app:boundary}. The limit of the boundary integral is $\delta\mathcal E$, provided we can interchange the limits above with differentiation with respect to $s$. This is justified in Appendix~\ref{app:boundary}.

By the estimates in Appendix~\ref{app:boundary}, the integrals over the puncture spheres and the outer sphere tend to zero. On each compact subinterval of an axis rod, the variations have traces $\delta v=\delta c_j$, $\delta\chi=\delta d_j$, and $\delta\psi=0$. The estimates in Appendix~\ref{app:boundary} then give
\begin{align}
 \delta_{J,Q}\mathcal{E}={}&-4\pi\sum_j\int_{\Gamma_j}\lim_{\rho\to0}
 \rho\left(e^{4u}\Theta_\rho\delta c_j
                         +e^{2u}\chi_\rho\delta d_j\right)\dd z\notag\\
 ={}&2\pi\sum_j\delta c_j[w]_{\Gamma_j}
                     +4\pi\sum_j\delta d_j[A_t]_{\Gamma_j}\notag\\
 ={}&8\pi\sum_i(\Omega_i\delta J_i+\Phi_i\delta Q_i).
 \label{eq:chargevariation}
\end{align}
For an oriented bounded axis rod $\Gamma_j=(a,b)$, we used the notation $[f]_{\Gamma_j}=f(b)-f(a)$. For a semi-infinite axis rod, take $b=+\infty$ or $a=-\infty$ and use the normalization of $w$ and $A_t$ at infinity. The last equality uses the jumps in Eq.~\eqref{eq:axisjumps} and $w,A_t\to0$ at both ends of the axis.

\subsection{Derivative with respect to a puncture position}

This computation was carried out in vacuum in our work on the mass--angular momentum inequality \cite{HKWXmass}. We present its Einstein--Maxwell extension here.

When a puncture moves, the truncated domain of integration moves with it. For a family of domains $G_s$ with boundary velocity $V$ and outward unit normal $n$, differentiation gives
\begin{equation}
 \left.\frac{d}{ds}\int_{G_s}f_s\,\dd^3x\right|_{s=0}
 =\int_{G_0}\dot f\,\dd^3x
  +\int_{\partial G_0}f\,V\cdot n\,\dd S.
 \label{eq:domainvariation}
\end{equation}
Here $f_s$ is the renormalized energy density. The first term is integrated by parts as above; the second accounts for motion of the boundary.

Let $p_i(s)=z_i(s)e_z$. Near this puncture define the map in centered coordinates by
\begin{equation}
 \begin{split}
 \widehat X_s(y)&=X_s(p_i(s)+y),\\
 \dot{\widehat X}(y)&=\dot X(p_i(0)+y)\\
 &\quad+\dot z_i\partial_zX(p_i(0)+y).
 \end{split}
 \label{eq:centeredvariation}
\end{equation}
Thus the derivative at a fixed spatial point splits into $-\dot z_i\partial_zX$ and the derivative $\dot{\widehat X}$ in centered coordinates. The local translation can be extended to a diffeomorphism equal to the identity outside a fixed neighborhood.

To write the resulting boundary terms, define, for $k,l\in\{\rho,z\}$,
\begin{equation}
 K_{kl}=U_kU_l+e^{4u}\Theta_k\Theta_l
                  +e^{2u}(\chi_k\chi_l+\psi_k\psi_l).
 \unreferencedlabel{eq:Ktensor}
 \notag
\end{equation}
Then $f=K_{\rho\rho}+K_{zz}$. The stress tensor associated with the renormalized harmonic map energy is
\begin{equation}
 T_{kl}=2K_{kl}-\delta_{kl}(K_{\rho\rho}+K_{zz}).
 \label{eq:stress}
\end{equation}
Let $\nu$ point away from the puncture, so the outward normal of the truncated exterior on the puncture sphere is $n=-\nu$. Substituting $-\dot z_i\partial_zX$ for $\delta X$ in the integrand of Eq.~\eqref{eq:energyboundary}, with $n=-\nu$, gives $2\dot z_iK_{\nu z}$. The moving-boundary term in Eq.~\eqref{eq:domainvariation} is $-\dot z_i f\nu_z$. Consequently, writing $\delta z_i=\dot z_i$, their combined contribution is
\begin{equation}
 \delta z_i\int_{S_r(p_i)}(2K_{\nu z}-f\nu_z)\,\dd S
 =\delta z_i\int_{S_r(p_i)}T_{\nu z}\,\dd S.
 \label{eq:shapederivative}
\end{equation}
Although the renormalized integrand contains explicit $\rho$ dependence, it has no explicit $z$ dependence. Using the harmonic map equations, this translation invariance gives $\partial_\rho(\rho T_{\rho z})+\partial_z(\rho T_{zz})=0$ away from the punctures. The Weyl quadratures read
\begin{equation}
 \alpha_\rho=\rho(K_{\rho\rho}-K_{zz}),\qquad
 \alpha_z=2\rho K_{\rho z}.
 \label{eq:weyl}
\end{equation}

Use polar coordinates $(r,\theta)$ centered at $p_i$, with $\theta=0$ pointing toward larger $z$. On a small sphere with normal directed away from the puncture, Eqs.~\eqref{eq:stress}--\eqref{eq:weyl} give
\begin{align}
 \int_{S_r(p_i)}T_{\nu z}\,\dd S
 &=2\pi r\int_0^\pi
       (\alpha_z\sin\theta-\alpha_\rho\cos\theta)\dd\theta\notag\\
 &=-2\pi\int_0^\pi\partial_\theta\alpha\,\dd\theta
   =2\pi(\alpha_{i+1}-\alpha_i).
 \label{eq:stressflux}
\end{align}

The remaining boundary term involves $\dot{\widehat X}$, the derivative in coordinates centered at the puncture. Its vanishing follows from the fixed-center estimates in (B2); the displacement contribution in Eq.~\eqref{eq:shapederivative} is retained. The interpolation of the identification away from the puncture introduces no additional term: on the truncated domain the Euler--Lagrange equations reduce the first variation to the boundary expression. Equation~\eqref{eq:stressflux} evaluates the remaining integral. Combining it with Eq.~\eqref{eq:chargevariation} yields
\begin{equation}
 \delta \mathcal{E}=2\pi\sum_i(\alpha_{i+1}-\alpha_i)\delta z_i
                 +8\pi\sum_i(\Omega_i\delta J_i+\Phi_i\delta Q_i).
 \label{eq:fullenergyvariation}
\end{equation}
Although the tangent map parameters do not appear explicitly in the flux in Eq.~\eqref{eq:stressflux}, in vacuum we have
\begin{equation}
 \alpha_{i+1}-\alpha_i=\log\frac{1+b_i}{1-b_i},
 \unreferencedlabel{eq:vacuumtangentparameter}
 \notag
\end{equation}
where $b_i$ is the tangent map parameter at the $i$-th puncture. Thus, in vacuum, Eq.~\eqref{eq:fullenergyvariation} agrees with the puncture variation formula in Ref.~\cite{HKWXmass}.

\subsection{The conical correction to ADM mass}

On the stationary maximal slice, let $g$ denote the induced spatial metric and $R_g$ its scalar curvature. The off-axis identity is
\begin{equation}
 e^{2(\alpha-U)}R_g=-2\Delta_2\alpha+4\Delta_3U-2|\nabla U|^2.
 \label{eq:scalarcurvature}
\end{equation}
The Einstein--Maxwell constraint and the stationary axisymmetric reconstruction give
\begin{equation}
 e^{2(\alpha-U)}R_g
 =2e^{4u}|\Theta|^2+2e^{2u}(|\nabla\chi|^2+|\nabla\psi|^2).
 \label{eq:constraint}
\end{equation}
Thus the integrand of $\mathcal{E}$ equals $2\Delta_3U-\Delta_2\alpha$. Integrating against $2\pi\rho\,\dd\rho\,\dd z$, using $U=-m/r+O(r^{-2})$, $\alpha=O(r^{-2})$, $\nabla\alpha=O(r^{-3})$, and the extremal endpoint estimates, gives
\begin{equation}
 \int\Delta_3U\,\dd^3x=4\pi m,\qquad
 \int\Delta_2\alpha\,\dd^3x=2\pi\sum_jL_j\alpha_j.
 \unreferencedlabel{eq:massboundaryintegrals}
 \notag
\end{equation}
It follows that
\begin{equation}
 m=\frac{\mathcal{E}}{8\pi}+\frac14\sum_jL_j\alpha_j.
 \label{eq:masscorrection}
\end{equation}
The logarithmic conical correction has an antecedent in force dependent mass inequalities \cite{Gabach}.

For punctures, $L_j=z_j-z_{j-1}$. Rearranging the sum and using telescoping cancellation gives
\begin{equation}
 \sum_i(\alpha_{i+1}-\alpha_i)\delta z_i
             =-\sum_j\alpha_j\delta L_j.
 \unreferencedlabel{eq:discrete}
 \notag
\end{equation}
Differentiating Eq.~\eqref{eq:masscorrection} yields:
\begin{align}
 \delta m
 &=\sum_i(\Omega_i\delta J_i+\Phi_i\delta Q_i)
                                      +\frac14\sum_jL_j\delta\alpha_j\notag\\
 &=\sum_i(\Omega_i\delta J_i+\Phi_i\delta Q_i)
                                      -\sum_j\lambda_j\delta\mu_j.
 \unreferencedlabel{eq:extremallaw}\notag
\end{align}
This proves the first law for an all extremal family.

\section{Nondegenerate horizons}
\label{sec:covariant}

We introduce a finite reduced harmonic map energy by subtracting a static reference solution with the same nondegenerate horizon rods, following Ref.~\cite{KSWPenrose}. The axis and puncture contributions were computed in Section~\ref{sec:energy}; here we compute the additional rod contributions. Part B gives an outline of the Iyer--Wald approach \cite{Wald,IyerWald} for completeness of the literature review; it is not used in this proof.

\subsection{Harmonic map approach}
\label{sec:harmonicnondeg}

\subsubsection{A static reference with the same rods}

For a nondegenerate horizon rod, $U=\log\rho+O(1)$, so the energy in Eq.~\eqref{eq:energy} diverges logarithmically. This prevents direct use of that integral, but not a harmonic map derivation of the first law. Following Khuri, Sokolowsky, and Weinstein \cite{KSWPenrose}, we subtract a static reference with the same nondegenerate rods. The subtraction removes the leading rod singularities before taking the first variation.

Use the static reference of Eq.~\eqref{eq:staticreference} and the reduced energy of Eq.~\eqref{eq:multirodenergy}. Set $\overline\beta=\overline\alpha-2\overline U$.
The reduced energy is the analog of Eq.~(2.19) in Ref.~\cite{KSWPenrose}, using the Bach--Weyl solution with the same horizon rods in place of the Schwarzschild reference solution.

\subsubsection{The reference-subtracted mass identity}

Our first aim is a mass formula for each solution. The scalar-curvature inequality used to obtain Eq.~(2.16) of Ref.~\cite{KSWPenrose} becomes an equality by Eq.~\eqref{eq:constraint}. That formula concerns a single horizon and regular exterior axes. We extend its boundary computation to several horizon rods and retain the contributions from the conical axis rods. With
\begin{equation}
 \mathcal B:=\frac14\sum_{i\in\mathcal N}
            \int_{z_i^-}^{z_i^+}\overline\beta(0,z)\,\dd z,
 \unreferencedlabel{eq:horizonmasscorrection}
 \notag
\end{equation}
the identity to be established is
\begin{align}
 m={}&m_*+\frac{\widehat{\mathcal E}}{8\pi}+\mathcal B\notag\\
    &+\frac14\sum_{j=2}^N L_j(\alpha_j-\alpha_{*,j}).
 \label{eq:multirodmass}
\end{align}

We first subtract the reference identities and then integrate by parts. Let $f$ and $\widehat f$ denote the integrands of $\mathcal E$ and $\widehat{\mathcal E}$. Equations~\eqref{eq:scalarcurvature} and \eqref{eq:constraint} give $f=2\Delta_3U-\Delta_2\alpha$, while the static reference satisfies $|\nabla U_*|^2=2\Delta_3U_*-\Delta_2\alpha_*$. Since $\Delta_3U_*=0$ off the axis,
\begin{align}
 \widehat f
 &=f-|\nabla U_*|^2-2\nabla\overline U\cdot\nabla U_*\notag\\
 &=2\Delta_3\overline U-\Delta_2\overline\alpha
       -2\nabla\overline U\cdot\nabla U_*\notag\\
 &=\operatorname{div}\overline{\boldsymbol W},
 \unreferencedlabel{eq:renormalizeddensity}\notag
\end{align}
where
\begin{equation}
 \overline{\boldsymbol W}
 =2\nabla\overline U-\nabla\overline\alpha
  +\frac{\overline\alpha}{\rho}\boldsymbol e_\rho
  -2\overline U\nabla U_*.
 \unreferencedlabel{eq:subtractedmassvector}
 \notag
\end{equation}
All derivatives are Euclidean. The last equality uses $\Delta_3\overline\alpha=\Delta_2\overline\alpha+\rho^{-1}\overline\alpha_\rho$ and $\operatorname{div}(\overline\alpha\boldsymbol e_\rho/\rho)=\overline\alpha_\rho/\rho$.

Write $G_{h,\varepsilon,\delta,R}$ for the region inside the sphere of radius $R$ after removing the cylinder $\rho\leq\delta$, balls of radius $h$ about the nondegenerate poles, and balls of radius $\varepsilon$ about the degenerate punctures. Choose disjoint balls and $0<\delta<\min(h,\varepsilon)$. We first take $\delta\to0$ with $h,\varepsilon,R$ fixed; the spherical boundary portions then tend to spheres. For $G=G_{h,\varepsilon,\delta,R}$ the divergence theorem gives
\begin{equation}
 \widehat{\mathcal E}_G
 =\int_{\partial G}\overline{\boldsymbol W}\cdot n\,\dd S,
 \label{eq:massintegrationbyparts}
\end{equation}
with $n$ outward from the truncated domain. We evaluate its outer-sphere, pole, puncture, and cylindrical boundary contributions in turn.

At infinity, Eq.~\eqref{eq:inputinfinity} gives
\begin{equation}
 \lim_{R\to\infty}\int_{S_R}
       \overline{\boldsymbol W}\cdot n\,\dd S=8\pi(m-m_*).
 \unreferencedlabel{eq:subtractedmassinfinity}
 \notag
\end{equation}

We use the spatial bounds in Eqs.~\eqref{eq:poleexpansion} and \eqref{eq:masspuncturebounds}.

On a spherical boundary portion $S_{r_p}$ centered on the axis, $\boldsymbol e_\rho\cdot n/\rho=\pm r_p^{-1}$. The explicit reference gives $\partial_{r_p}U_*=O(r_p^{-1})$ at a rod endpoint and is smooth at a degenerate puncture. Consequently,
\begin{equation}
 \int_{S_{r_p}}|\overline{\boldsymbol W}\cdot n|\,\dd S
 \leq
 \begin{cases}
 Cr_p, &\text{at a pole},\\
 C\big(r_p+r_p^2|\log r_p|\big), &\text{at a puncture}.
 \end{cases}
 \unreferencedlabel{eq:masscaplimits}
 \notag
\end{equation}
Thus the pole and puncture integrals tend to zero in the mass identity.

On the cylindrical boundary, $n=-\boldsymbol e_\rho$ and $\dd S=\rho\,\dd\phi\,\dd z$. Its contribution is
\begin{equation}
 -2\pi\int_{\rho=\delta}\big[2\rho\overline U_\rho+\overline\alpha-\rho\overline\alpha_\rho-2\rho\overline U(U_*)_\rho\big]\,\dd z,
 \label{eq:cylindermassintegral}
\end{equation}
where the integral is over the cylinder outside the pole and puncture balls. We separate the portions approaching axis rods from those approaching horizon rods.

On a compact subinterval of an axis rod $\Gamma_j$, we have $\rho\overline U_\rho,\rho\overline\alpha_\rho\to0$ and $\rho(U_*)_\rho\to0$. The integrand in Eq.~\eqref{eq:cylindermassintegral} therefore tends to $\alpha_j-\alpha_{*,j}$.

On a compact subinterval of a horizon rod, we again have $\rho\overline U_\rho,\rho\overline\alpha_\rho\to0$, but $\rho(U_*)_\rho\to1$. The integrand in Eq.~\eqref{eq:cylindermassintegral} therefore tends to
\begin{equation}
 \overline\alpha(0,z)-2\overline U(0,z)
 =\overline\beta(0,z).
 \unreferencedlabel{eq:subtractedboundary}
 \notag
\end{equation}

Equation~\eqref{eq:poleexpansion} provides a uniform bound for the integrand in Eq.~\eqref{eq:cylindermassintegral}. Near a puncture, the reference is smooth and axisymmetric, so $(U_*)_\rho=O(\rho)$; Eq.~\eqref{eq:masspuncturebounds} again gives a uniform bound. Thus the integrals over portions of length $h$ near these points are $O(h)$, uniformly for smaller cylinder radius. The semi-infinite axis rods contribute zero by the asymptotic normalization.

Let $\delta\to0$ with $h,\varepsilon,R$ fixed, then $h\to0$, then $\varepsilon\to0$, and finally $R\to\infty$. Since the reduced energy density is nonnegative, monotone convergence shows that the limit of the left-hand side of Eq.~\eqref{eq:massintegrationbyparts} equals $\widehat{\mathcal E}$. The boundary limits therefore give
\begin{align}
 \widehat{\mathcal E}={}&8\pi(m-m_*)
      -2\pi\sum_{i\in\mathcal N}
          \int_{z_i^-}^{z_i^+}\overline\beta(0,z)\,\dd z\notag\\
      &-2\pi\sum_{j=2}^N L_j(\alpha_j-\alpha_{*,j}),
 \unreferencedlabel{eq:multirodcutoff}\notag
\end{align}
which proves Eq.~\eqref{eq:multirodmass} and, in particular, $\widehat{\mathcal E}<\infty$.

\subsubsection{Variation of the energy and horizon terms}

We now vary the mass identity along a curve in the parameter space.

On a fixed truncated domain $G=G_{h,\varepsilon,\delta,R}$, subtraction of the static reference changes only the $U$ term:
\begin{equation}
 \widehat{\mathcal{E}}_G-\mathcal{E}_G
 =\int_G\left(|\nabla U_*|^2-2\nabla U\cdot\nabla U_*\right)\dd^3x.
 \label{eq:energydifference}
\end{equation}
The variation of Eq.~\eqref{eq:energydifference} on a fixed domain is
\begin{align}
 \delta\widehat{\mathcal{E}}_G-\delta \mathcal{E}_G
 =-2\int_G\big(&\nabla U_*\cdot\nabla\delta U\notag\\
 &+\nabla\overline U\cdot\nabla\delta U_*\big)\dd^3x.
 \unreferencedlabel{eq:differencevariation}\notag
\end{align}
If the domain is also moving, we must add
\begin{equation}
 \int_{\partial G}
 \left(|\nabla U_*|^2-2\nabla U\cdot\nabla U_*\right)
 (V\cdot n)\,\dd S,
 \unreferencedlabel{eq:energydifferencemotion}
 \notag
\end{equation}
where $V$ is the velocity of the boundary along the parameter variation and $n$ is the outward unit normal. The reference subtraction leaves the twist and electromagnetic terms unchanged. The terms arising from variations of the axis potential data and of the puncture positions are therefore the same as in Sec.~\ref{sec:energy}.

To compute the horizon contribution, first express $\mathcal B$ in terms of rod lengths and surface gravities. On the interior of a nondegenerate horizon rod, the Weyl metric gives the area density and surface gravity relation
\begin{equation}
 \dd A_i=\frac1{\kappa_i}\,\dd z\,\dd\phi,
 \qquad A_i=\frac{2\pi\ell_i}{\kappa_i}.
 \label{eq:rodarea}
\end{equation}
Here $\dd$ denotes the area element; variations are denoted by $\delta$. Indeed, the area density is $\lim_{\rho\to0}\rho e^{\alpha-2U}$, the reciprocal of $\kappa_i$ in the stated normalization. Comparing physical and reference horizons therefore gives
\begin{equation}
 \overline\beta|_{\mathcal H_i}
   =\log\frac{\kappa_{*,i}}{\kappa_i},\qquad
 \mathcal B=\frac14\sum_{i\in\mathcal N}
       \ell_i\log\frac{\kappa_{*,i}}{\kappa_i}.
 \label{eq:referencebeta}
\end{equation}
The surface gravities are constant on each horizon. The bounds in Eq.~\eqref{eq:poleexpansion} ensure that the horizon integrals converge at their endpoints.

Differentiating this finite sum, with $\delta\ell_i=\delta z_i^+-\delta z_i^-$, gives
\begin{align}
 \delta\mathcal B=\frac14\sum_{i\in\mathcal N}\bigg[
 &\log\frac{\kappa_{*,i}}{\kappa_i}\,\delta\ell_i
 +\ell_i\delta\log\kappa_{*,i}\notag\\
 &-\ell_i\delta\log\kappa_i\bigg].
 \unreferencedlabel{eq:Bvariation}\notag
\end{align}
From Eq.~\eqref{eq:rodarea},
\begin{equation}
 \frac{\kappa_i}{8\pi}\delta A_i
 =\frac14\delta\ell_i-\frac{\ell_i}{4}\delta\log\kappa_i.
 \unreferencedlabel{eq:rodareavariation}
 \notag
\end{equation}
Consequently,
\begin{align}
 \delta\mathcal B={}&\sum_{i\in\mathcal N}\frac{\kappa_i}{8\pi}\delta A_i\notag\\
 &+\frac14\sum_{i\in\mathcal N}\left[
 \left(\log\frac{\kappa_{*,i}}{\kappa_i}-1\right)\delta\ell_i
 +\ell_i\delta\log\kappa_{*,i}\right].
 \label{eq:Barea}
\end{align}

\subsubsection{Direct variation of the reduced energy}
\label{sec:directreferencevariation}

Write $a_i=z_i^-$, $b_i=z_i^+$ for $i\in\mathcal N$. The static potential has the line-integral representation
\begin{equation}
 U_*=-\frac12\sum_{i\in\mathcal N}\int_{a_i}^{b_i}
                  \frac{\dd z'}{|x-z'e_z|}.
 \unreferencedlabel{eq:referencepotentialintegral}
 \notag
\end{equation}
Consequently, with $r_{a_i}=|x-a_ie_z|$ and $r_{b_i}=|x-b_ie_z|$,
\begin{equation}
 \delta U_* =\frac12\sum_{i\in\mathcal N}
       \left(\frac{\delta a_i}{r_{a_i}}-\frac{\delta b_i}{r_{b_i}}\right),
 \qquad \delta m_*=\frac12\sum_{i\in\mathcal N}\delta\ell_i.
 \label{eq:referenceendpointvariation}
\end{equation}
This variation is taken at fixed $x$; it includes the singular contribution of each moving endpoint.

To differentiate the energy, put
$\overline P=2e^{4u}|\Theta|^2+e^{2u}(|\nabla\chi|^2+|\nabla\psi|^2)$.
The harmonic map equation gives $\Delta\overline U=\overline P$ off the axis. Integrating by parts the terms containing gradients of $\delta U$, $\delta v$, $\delta\chi$, and $\delta\psi$, and using the harmonic map equations, leaves the bulk term $2\int_G \overline P\delta U_*\,\dd^3x$. Since $\Delta\delta U_*=0$ on $G$, Green's identity converts it to a boundary integral. Combining it with the variation of $|\nabla\overline U|^2$ gives
\begin{align}
 \delta\widehat{\mathcal E}_G
 ={}&2\int_{\partial G}\big[\delta U\,\partial_n\overline U
       -\overline U\,\partial_n\delta U_*+\mathfrak j_n\big]\dd S\notag\\
 &+\int_{\partial G}\widehat f\,V\cdot n\,\dd S,
 \unreferencedlabel{eq:directreferenceboundary}\notag
\end{align}
where $\mathfrak j_n$ is defined in Eq.~\eqref{eq:potentialvariationcurrent} and $\widehat f=|\nabla\overline U|^2+P$, where we recall that $P$ is defined in Eq.~\eqref{eq:potentialdensity}.

At a pole $p$, let $S_{r_p}$ be the sphere of radius $r_p=h$, obtained after $\delta\to0$ with the pole ball fixed. The pole estimates in (B1)--(B2) and Eq.~\eqref{eq:polecurrentbound} imply
\begin{align}
 \int_{S_{r_p}}|\delta U\,\partial_n\overline U|\dd S
       &\leq C(r_p^\gamma+r_p^{2\gamma}),\notag\\
 \int_{S_{r_p}}\widehat f\,|V\cdot n|\dd S
       &\leq C(r_p^{2\gamma}+r_p),\notag\\
 \int_{S_{r_p}}|\mathfrak j_n|\dd S&\leq Cr_p^\eta.
 \unreferencedlabel{eq:variationpoleerrors}\notag
\end{align}
Here $r_p<1$ and $n$ points into the excluded ball. To see the first estimate, use $\delta\overline U=D_s\overline U-\dot p\,\partial_z\overline U$, since $D_s$ differentiates at fixed coordinates $y=x-p(s)e_z$, which track the moving pole, together with the $r_p^{-1}$ singularity in Eq.~\eqref{eq:referenceendpointvariation}. The singular part of $\delta U_*$ is $\delta a_i/(2r_{a_i})$ at $a_i$ and $-\delta b_i/(2r_{b_i})$ at $b_i$. Since $\int_{S_{r_p}}\partial_n(r_p^{-1})\,\dd S=4\pi$, the sphere gives, as $h\to0$,
\begin{equation}
 -2\int_{S_{r_p}}\overline U\,\partial_n\delta U_*\,\dd S
 \longrightarrow
 \begin{cases}
 -4\pi\overline U(a_i)\delta a_i,&p=a_i,\\
  4\pi\overline U(b_i)\delta b_i,&p=b_i.
 \end{cases}
 \label{eq:referencepoleflux}
\end{equation}
Replacing $\overline U$ by $\overline U(p)$ in the singular reference flux introduces an $O(h^\gamma)$ error; the smooth reference remainder tends to zero as $h\to0$.

The full boundary current on each compact subinterval of a horizon
cylinder tends to zero by Appendix~\ref{app:linearconditions}.
The endpoint-tail condition there removes the remaining short-cylinder
contributions after the cylinder limit; the sphere residues in
Eq.~\eqref{eq:referencepoleflux} are retained. The outer boundary
contribution vanishes by the outer-sphere hypotheses in
Appendix~\ref{app:linearconditions}. On axis rods the potential-current calculation of Sec.~\ref{sec:energy} gives $8\pi\sum_i(\Omega_i\delta J_i+\Phi_i\delta Q_i)$. At each degenerate puncture the reference is smooth, and its subtraction changes none of the surviving translation flux in Eq.~\eqref{eq:stressflux}.

Set $\overline\beta_i=\log(\kappa_{*,i}/\kappa_i)$. Continuity of the fields at the poles and Eq.~\eqref{eq:referencebeta} give
\begin{align}
 2\overline U(a_i)&=\alpha_i-\alpha_{*,i}-\overline\beta_i,\notag\\
 2\overline U(b_i)&=\alpha_{i+1}-\alpha_{*,i+1}-\overline\beta_i.
 \unreferencedlabel{eq:referencepoletraces}\notag
\end{align}
For a degenerate component $z_i$, the reference constants agree on its two adjacent axis rods. Thus the pole terms in Eq.~\eqref{eq:referencepoleflux}, together with the degenerate displacement terms, form a telescoping sum. Since $\delta L_j=\delta z_j^--\delta z_{j-1}^+$, we obtain
\begin{align}
 \frac{\delta\widehat{\mathcal E}}{8\pi}
 ={}&\sum_i(\Omega_i\delta J_i+\Phi_i\delta Q_i)
       -\frac14\sum_{i\in\mathcal N}\overline\beta_i\delta\ell_i\notag\\
 &-\frac14\sum_{j=2}^N(\alpha_j-\alpha_{*,j})\delta L_j.
 \label{eq:directenergyvariation}
\end{align}

\subsubsection{Static reference variations and the first law}

The explicit static solution satisfies the differential identity
\begin{equation}
 \sum_{i\in\mathcal N}\delta\ell_i
 +\sum_{i\in\mathcal N}\ell_i\delta\log\kappa_{*,i}
 =\sum_{j=2}^N L_j\delta\alpha_{*,j}.
 \label{eq:staticreferenceidentity}
\end{equation}
We verify it from the rod formulas. In this paragraph only, number the nondegenerate reference rods consecutively from $1$ to $n=|\mathcal N|$. For $p<q$, define
\begin{equation}
 C_{pq}=\frac{(a_q-b_p)(b_q-a_p)}{(a_q-a_p)(b_q-b_p)}.
 \unreferencedlabel{eq:referencecrossratio}
 \notag
\end{equation}
On the gap between consecutive reference rods $k,k+1$, the static constant is $\sum_{p\leq k<q}\log C_{pq}$. The surface gravities obey
\begin{align}
 \log(2\ell_i\kappa_{*,i})
 ={}&\sum_{q>i}\log\frac{a_q-a_i}{b_q-a_i}
       +\sum_{p<i}\log\frac{b_i-b_p}{b_i-a_p}\notag\\
 &-\sum_{p<i<q}\log C_{pq}.
 \label{eq:statickappaformula}
\end{align}
For a fixed pair $p<q$, put $x=\ell_p$, $y=\ell_q$, and $g=a_q-b_p$. Direct differentiation gives
\begin{align}
 &x\,\delta\log\frac{g+x}{g+x+y}
 +y\,\delta\log\frac{g+y}{g+x+y}\notag\\
 &\hspace{1em}=g\,\delta\log\frac{g(g+x+y)}{(g+x)(g+y)}.
 \label{eq:referencepairidentity}
\end{align}
When Eq.~\eqref{eq:statickappaformula} is multiplied by $\ell_i$ after differentiation and summed over $i$, each pair contributes the right-hand side of Eq.~\eqref{eq:referencepairidentity} minus $\sum_{p<i<q}\ell_i\delta\log C_{pq}$. The resulting coefficient is the total length of the intervening gaps. Summing over pairs proves Eq.~\eqref{eq:staticreferenceidentity}. Any degenerate punctures merely subdivide these gaps; their adjacent reference constants agree.

Combining Eqs.~\eqref{eq:Barea}, \eqref{eq:referenceendpointvariation}, and \eqref{eq:staticreferenceidentity} gives
\begin{align}
 \delta m_*+\delta\mathcal B
 ={}&\sum_{i\in\mathcal N}\frac{\kappa_i}{8\pi}\delta A_i
       +\frac14\sum_{i\in\mathcal N}\overline\beta_i\delta\ell_i\notag\\
 &+\frac14\sum_{j=2}^N L_j\delta\alpha_{*,j}.
 \unreferencedlabel{eq:referencehorizoncancellation}\notag
\end{align}
Substitution in the directly computed energy variation \eqref{eq:directenergyvariation} yields
\begin{align}
 \frac{\delta\widehat{\mathcal E}}{8\pi}
 ={}&\sum_{i\in\mathcal N}\frac{\kappa_i}{8\pi}\delta A_i
       -\delta m_*-\delta\mathcal B\notag\\
 &+\sum_i(\Omega_i\delta J_i+\Phi_i\delta Q_i)\notag\\
 &-\frac14\sum_{j=2}^N(\alpha_j-\alpha_{*,j})\delta L_j
       +\frac14\sum_{j=2}^N L_j\delta\alpha_{*,j}.
 \unreferencedlabel{eq:renormalizedvariationtarget}\notag
\end{align}
Differentiating Eq.~\eqref{eq:multirodmass} now cancels the reference terms and the terms proportional to $\delta L_j$, leaving
\begin{align}
 \delta m={}&\sum_{i\in\mathcal N}\frac{\kappa_i}{8\pi}\delta A_i\notag\\
 &+\sum_i(\Omega_i\delta J_i+\Phi_i\delta Q_i)
        +\frac14\sum_{j=2}^N L_j\delta\alpha_j.
 \unreferencedlabel{eq:harmonicfirstlawtarget}\notag
\end{align}
Since $L_j\delta\alpha_j/4=-\lambda_j\delta\mu_j$, this proves the first law by the harmonic map approach under (B1)--(B2), with the analytic scope specified in Sec.~\ref{sec:input}.

\subsection{Outline of the Iyer--Wald approach}
\label{sec:iyerwald}
\label{sec:completefirstlaw}

For completeness of the literature review, we outline the boundary-identity approach of Wald and Iyer--Wald \cite{Wald,IyerWald}, with the electromagnetic horizon treatment of Gao \cite{Gao}. This outline is not used in the harmonic map proof of the first law above. For the action $\int\boldsymbol L$, let $\boldsymbol\vartheta$ be its first variation boundary three-form and $\boldsymbol Q_\xi$ its Noether charge two-form. Along a family of solutions, with the symmetry field $\xi$ held fixed,
\begin{equation}
 \boldsymbol k_\xi=\delta\boldsymbol Q_\xi-\iota_\xi\boldsymbol\vartheta,
 \qquad \dd\boldsymbol k_\xi=0.
 \unreferencedlabel{eq:kform}
 \notag
\end{equation}
With the background generator $\xi_i=\tau+\Omega_i\eta$ and its
background coefficient $\Omega_i$ held fixed in the charge variation,
\begin{equation*}
 \boldsymbol k_\tau=\boldsymbol k_{\xi_i}-\Omega_i\boldsymbol k_\eta,
 \qquad \int_{H_i}\boldsymbol k_\eta=-\delta J_i.
\end{equation*}
The standard infinity and horizon integrals give $\delta m$ and
\begin{equation}
 \begin{aligned}
 \int_{H_i}\boldsymbol k_\tau
 ={}&\Omega_i\delta J_i+\Phi_i\delta Q_i\\
 &+\begin{cases}
 \dfrac{\kappa_i}{8\pi}\delta A_i,&i\in\mathcal N,\\
 0,&i\notin\mathcal N.
 \end{cases}
 \end{aligned}
 \label{eq:fullhorizon}
\end{equation}
The additional step for our configurations is to excise tubes around the conical struts and small neighborhoods of their intersections with the horizons. With each tube oriented outward from the removed region, the conical metric expansion gives the limiting contribution $-\lambda_j\delta\mu_j$. There is no additional Maxwell tube term in the regular, uncharged-strut sector. The additional boundary conditions in Appendix~\ref{app:covariantconditions} control the connecting caps and the removal of the cutoffs. Stokes' theorem then gives Eq.~\eqref{eq:firstlaw}. Degenerate horizon cross-sections are retained; they are not treated as vanishing caps. The local horizon and cap conditions are recorded in the appendices.

\section{The corrected mass function and its derivatives}
\label{sec:potential}

Recall the corrected mass function $\cE$ defined in Eq.~\eqref{eq:potentialdefinition}.
On the all-extremal family this is $\mathcal{E}/(8\pi)$ by Eq.~\eqref{eq:masscorrection}. Since $-\lambda_j\dd\mu_j=(L_j/4)\dd\alpha_j$, the first law is equivalent to
\begin{align}
 \dd\cE={}&\sum_{i\in\mathcal N}\frac{\kappa_i}{8\pi}\dd A_i\notag\\
 &+\sum_i(\Omega_i\dd J_i+\Phi_i\dd Q_i)
                         -\frac14\sum_j\alpha_j\dd L_j.
 \label{eq:potentiallaw}
\end{align}

On an all-degenerate family with independent coordinates $(L,J,Q)$, Eq.~\eqref{eq:potentiallaw} gives
\begin{equation}
 \partial_{J_i}\cE=\Omega_i,\qquad
 \partial_{Q_i}\cE=\Phi_i,\qquad
 \partial_{L_j}\cE=-\frac14\alpha_j.
 \label{eq:potentialpartials}
\end{equation}
Whenever $C^2$ regularity permits commuting derivatives,
\begin{align}
 \partial_{L_j}\Omega_i&=-\tfrac14\partial_{J_i}\alpha_j,&
 \partial_{L_j}\Phi_i&=-\tfrac14\partial_{Q_i}\alpha_j,\notag\\
 \partial_{Q_k}\Omega_i&=\partial_{J_i}\Phi_k,&
 \partial_{J_k}\Omega_i&=\partial_{J_i}\Omega_k.
 \unreferencedlabel{eq:reciprocity}\notag
\end{align}
All remaining coordinates are held fixed. These mixed-derivative identities are useful consistency conditions for numerical stationary solutions, but their symmetry alone supplies no sign information.

The scaling variation supplies a further check. Under a common dilation by $s>0$, the family satisfies
\begin{align}
 m(s)&=sm,& A_i(s)&=s^2A_i,& J_i(s)&=s^2J_i,\notag\\
 Q_i(s)&=sQ_i,& L_j(s)&=sL_j,& \mu_j(s)&=\mu_j.
 \unreferencedlabel{eq:scalingcurve}\notag
\end{align}
At $s=1$ its tangent therefore has $\dot m=m$, $\dot A_i=2A_i$, $\dot J_i=2J_i$, $\dot Q_i=Q_i$, and $\dot\mu_j=0$. Evaluating the first-law one-form on this tangent gives
\begin{equation}
 m=\sum_i\left(\frac{\kappa_i A_i}{4\pi}
                             +2\Omega_iJ_i+\Phi_iQ_i\right).
 \unreferencedlabel{eq:smarr}
 \notag
\end{equation}
There is no explicit strut term because the force is dimensionless.

For the Bach--Weyl two body solution, the identity can be checked directly by varying the two masses and their separation; see Appendix~\ref{sec:binarycheck}.

\section{Spin-Hessian comparison near the Majumdar--Papapetrou family}
\label{sec:response}

\label{sec:hessianinput}

For the results of this section we additionally require $\cE$ to be $C^2$ in $J$ near $J=0$, at fixed $(L,Q)$, and the fields and spatial gradients to be differentiable along spin variations away from the singular set, with preserved differentiated axis traces. We assume continuity of $\partial_{L_j}\cE$ along the comparison paths. Each nearby configuration must satisfy the hypotheses of Theorem~1.1 and Definition~2.1 of Ref.~\cite{HKWXev}, including its axis analyticity and puncture-profile conditions. These hypotheses are needed throughout a common spin neighborhood, not only at the Majumdar--Papapetrou solution.

\subsection{Majumdar--Papapetrou as the reference family}

Fix $Q_i>0$ and ordered, distinct punctures. The electric Majumdar--Papapetrou fields are
\begin{align}
 H(x)&=1+\sum_i\frac{Q_i}{|x-p_i|},\notag\\
 \dd s^2&=-H^{-2}\dd t^2+H^2\dd x^2,
 \qquad \cA_{\mathrm{ext}}=(H^{-1}-1)\dd t.
 \unreferencedlabel{eq:MP}\notag
\end{align}
The potential $\cA_{\mathrm{ext}}$ is an exterior representative.
Near the $i$-th center, write
\begin{gather*}
 r_i=|x-p_i|,\qquad H=\frac{Q_i}{r_i}+h_i,\\
 h_{i0}=1+\sum_{k\ne i}\frac{Q_k}{|p_i-p_k|}.
\end{gather*}
For fixed $r_0>0$, introduce
\begin{align*}
 V_i&=t-\frac{Q_i^2}{r_i}+2Q_i h_{i0}\log(r_i/r_0),\\
 \Lambda_i&=-\frac{Q_i^2}{r_i}
       +Q_i(2h_{i0}-1)\log(r_i/r_0).
\end{align*}
In the coordinates $(V_i,r_i,\omega_i)$, which are regular across the
future horizon,
\begin{align*}
 \cA_{\mathrm{ext}}-\dd\Lambda_i
 &=\left(\frac{r_i}{Q_i+r_i h_i}-1\right)\dd V_i\\
 &\quad+\frac{Q_i(h_i-2h_{i0})}{Q_i+r_i h_i}\,\dd r_i
\end{align*}
is regular at $r_i=0$. Choose smooth stationary, axisymmetric cutoffs
$\zeta_i$ with disjoint supports, equal to one near the respective centers
and zero near infinity. Then
\begin{equation*}
 \cA=\cA_{\mathrm{ext}}-\dd\!\left(\sum_i\zeta_i\Lambda_i\right)
\end{equation*}
is regular across all future horizons and satisfies
\begin{equation*}
 \dd\cA=F,\qquad \cA(\tau)=H^{-1}-1,\qquad \cA(\eta)=0.
\end{equation*}
Thus the normalization at infinity and the values $\Phi_i=1$ are unchanged.
Regularity of the nearby rotating family remains among our hypotheses.

The MP solutions have $\alpha=0$, $J_i=\Omega_i=0$, $\Phi_i=1$, and $m=\cE=\sum_iQ_i$, independently of the positions \cite{Majumdar,Papapetrou,ChruscielNadirashvili}. Thus $\cE(L,0,Q)$ is independent of the gaps.

Consider a neighborhood of this family with all horizons still degenerate and $(L,J,Q)$ independent local data. The prescribed potential jumps allow independent $J_i$ and $Q_i$ after fixing the gauges. We use the existence theory described in Sec.~\ref{sec:input} and retain its additional differentiability and boundary hypotheses. We also assume every configuration in the comparison family satisfies the hypotheses of Theorem~1.1 of Ref.~\cite{HKWXev}. These include the prescribed black hole asymptotics, uniform puncture Weyl-profile control, and the analyticity on axis rods of the renormalized potentials. That result gives
\begin{equation}
 \alpha_j\leq0\quad\text{for every finite gap},
 \label{eq:forcesign}
\end{equation}
with equality only for Majumdar--Papapetrou solutions and strict negativity on every finite rod otherwise. This is Theorem~1.1 of Ref.~\cite{HKWXev}, applied in our positive electric-charge sector. Its hypotheses explicitly require that the corresponding renormalized harmonic map be analytic on the axis rods, in addition to the other asymptotic conditions in Definition~2.1 of that reference. In our convention, negative $\alpha_j$ corresponds to positive $\mu_j=(e^{-\alpha_j}-1)/4$.

\subsection{Square completion and the spin Hessian}
\label{sec:squarecompletion}

Define a divergence-free Euclidean vector field from the electric axial potential:
\begin{equation}
 \Eflat=\frac1\rho\left(\chi_z\boldsymbol e_\rho
                                -\chi_\rho\boldsymbol e_z\right).
 \unreferencedlabel{eq:flatE}
 \notag
\end{equation}
Its flux around puncture $i$ is $4\pi Q_i$, and
$e^{2U}|\Eflat|^2=e^{2u}|\nabla\chi|^2$.
For each solution in the family, we assume that the flux of $e^U\Eflat$ across the inner boundary tends to zero as the axis and puncture cutoffs are removed, and that its flux at infinity is $4\pi\sum_iQ_i$. Completing the square in Eq.~\eqref{eq:energy} then yields
\begin{align}
 \cE=\sum_iQ_i+\frac1{8\pi}\int_{\R^3}\big[&
 |\nabla U-e^U\Eflat|^2+e^{4u}|\Theta|^2\notag\\
 &+e^{2u}|\nabla\psi|^2\big]\dd^3x.
 \label{eq:square}
\end{align}
Indeed, the cross term is $2\operatorname{div}(e^U\Eflat)$. Its integral contributes $8\pi\sum_iQ_i$ from the boundary at infinity. At each puncture, the bounds $e^U=O(r_i)$ and $\Eflat\cdot n=O(r_i^{-2})$, uniform in the angular variables, imply that the flux of $e^U\Eflat$ through a sphere of radius $r_i$ is $O(r_i)$ and hence tends to zero. Regularity on the axis makes the flux through the cylindrical boundary tend to zero. This bounds the corrected mass $\cE$ from below by $\sum_iQ_i$. It does not give the same bound for $m$, since $m=\cE+\frac14\sum_jL_j\alpha_j$ and $\alpha_j\leq0$.

At the reference fields, $\psi=0$, $\Theta=0$, and $\Eflat=-\nabla H$, so every square vanishes. Define the Hessian with respect to angular momenta, holding $(L,Q)$ fixed
\begin{equation}
 \HH_{ik}(L,Q):=\left.\frac{\partial^2\cE}{\partial J_i\partial J_k}\right|_{J=0}
              =\left.\frac{\partial\Omega_i}{\partial J_k}\right|_{J=0}.
 \unreferencedlabel{eq:Hdefinition}
 \notag
\end{equation}
The second equality follows from the first law. The Hessian is symmetric, and
\begin{equation}
 \boldsymbol\Omega=\HH\boldsymbol J+o(|J|),\qquad
 \cE=\sum_iQ_i+\tfrac12\boldsymbol J^T\HH\boldsymbol J+o(|J|^2).
 \label{eq:responseexpansion}
\end{equation}

\begin{proposition}[Positive spin Hessian] Under (B1)--(B2), the spin differentiability hypotheses in Sec.~\ref{sec:hessianinput}, the boundary flux limits stated above, and the differentiated axis conditions in Eq.~\eqref{eq:linearizedtraces}, the matrix $\HH(L,Q)$ is positive definite. \end{proposition}

Fix $\mathbf a\in\R^N$ and set $J=\varepsilon\mathbf a$. A dot denotes differentiation at $\varepsilon=0$. The prescribed axis values give
\begin{equation}
 \dot\psi|_{\Gamma_k}=0,\qquad
 \dot v|_{\Gamma_{i+1}}-\dot v|_{\Gamma_i}=4a_i.
 \label{eq:linearizedtraces}
\end{equation}
Here we assume that parameter differentiation commutes with taking the axis limits. This assumption and the boundary flux limits above require estimates for the constructed family beyond differentiability away from the axis. Write
\begin{equation}
 \boldsymbol B_\varepsilon
 =\nabla U_\varepsilon-e^{U_\varepsilon}\Eflat_\varepsilon.
 \notag
\end{equation}
At the reference MP solution, $\boldsymbol B_0=0$, $\Theta_0=0$, and $\nabla\psi_0=0$. Thus Eq.~\eqref{eq:square} gives, for $\varepsilon\neq0$,
\begin{equation}
 \frac{8\pi(\cE_\varepsilon-\cE_0)}{\varepsilon^2}
   =\int_{\R^3}F_\varepsilon\,\dd^3x,
 \label{eq:fatouquotient}
\end{equation}
where $\cE_\varepsilon=\cE(L,\varepsilon\mathbf a,Q)$ and
\begin{align}
 F_\varepsilon={}&\left|\frac{\boldsymbol B_\varepsilon}{\varepsilon}\right|^2
 +e^{4u_\varepsilon}\left|\frac{\Theta_\varepsilon}{\varepsilon}\right|^2\notag\\
 &+e^{2u_\varepsilon}\left|\frac{\nabla\psi_\varepsilon}{\varepsilon}\right|^2\geq0.
 \unreferencedlabel{eq:fatouintegrand}\notag
\end{align}
Parameter differentiability away from the singular set gives almost-everywhere convergence to
\begin{equation}
 F_*:=|\dot{\boldsymbol B}|^2+e^{4u_0}|\dot\Theta|^2
                           +e^{2u_0}|\nabla\dot\psi|^2.
 \notag
\end{equation}
Fatou's lemma gives
\begin{equation}
 \int_{\R^3}F_*\,\dd^3x
 \leq\liminf_{\varepsilon\to0}\int_{\R^3}F_\varepsilon\,\dd^3x.
 \label{eq:fatoustep}
\end{equation}
The assumed $C^2$ dependence of $\cE$ and its vanishing first spin derivative at MP give
\begin{equation}
 \cE_\varepsilon-\cE_0
 =\tfrac12\varepsilon^2\mathbf a^T\HH\mathbf a+o(\varepsilon^2).
 \notag
\end{equation}
Substituting this expansion into Eq.~\eqref{eq:fatouquotient} and using Eq.~\eqref{eq:fatoustep} yields
\begin{equation}
 \mathbf a^T\HH\mathbf a\geq\frac1{4\pi}\int_{\R^3}\left[
 |\dot{\boldsymbol B}|^2+e^{4u_0}|\dot\Theta|^2
                         +e^{2u_0}|\nabla\dot\psi|^2\right]\dd^3x.
 \label{eq:Hpositive}
\end{equation}
If $\mathbf a^T\HH\mathbf a=0$, the right-hand side vanishes, so $\dot\psi$ is constant and its zero axis trace makes it zero. At Majumdar--Papapetrou, $\dot\Theta=\dd\dot v+\chi_0\dd\dot\psi-\dot\psi\dd\chi_0=\dd\dot v$. Thus $\dot v$ is constant as well. Its differentiated axis jumps are $4a_i$, forcing every $a_i=0$. Hence
\begin{equation}
 \HH(L,Q)>0.
 \label{eq:Hpd}
\end{equation}

\subsection{Hessian comparison and matrix corollaries}

\begin{theorem}[Hessian comparison]
Fix $Q_i>0$ and a piecewise $C^1$, coordinatewise nondecreasing path $L(s)$, $0\leq s\leq1$, from $L$ to $L'$. Suppose a common open neighborhood of $J=0$ along the path consists of all-degenerate solutions satisfying the hypotheses of Theorem~1 and the square-completion and spin differentiability hypotheses of Proposition~2. Assume $\partial_{L_j}\cE$ is continuous along the path and every solution in this neighborhood belongs to the asymptotic class of Definition~2.1 and satisfies Theorem~1.1 of Ref.~\cite{HKWXev}. Then
\begin{equation}
 \HH(L',Q)-\HH(L,Q)\geq0.
 \label{eq:Horder}
\end{equation}
\end{theorem}

To prove it, hold $J,Q$ fixed and integrate Eq.~\eqref{eq:potentialpartials} along the path $L(s)$:
\begin{equation}
 \begin{aligned}
 D(J)&:=\cE(L',J,Q)-\cE(L,J,Q)\\
     &=-\frac14\int_0^1\sum_j\alpha_j(L(s),J,Q)
                              \frac{\dd L_j}{\dd s}\,\dd s\geq0.
 \end{aligned}
 \unreferencedlabel{eq:comparisonproof}
 \notag
\end{equation}
The inequality follows from $\alpha_j\leq0$ and $\dd L_j/\dd s\geq0$. At $J=0$, both endpoint configurations are Majumdar--Papapetrou, so $D(0)=0$. Thus $D$ has a minimum at $J=0$, and its Hessian $\HH(L',Q)-\HH(L,Q)$ is positive semidefinite.

For each fixed angular momentum direction $\mathbf a$, the second derivative of $\cE(L,\varepsilon\mathbf a,Q)$ at $\varepsilon=0$ is $\mathbf a^T\HH\mathbf a$, which is nondecreasing along the path. The corresponding angular velocity satisfies $\mathbf a^T\boldsymbol\Omega=\varepsilon\mathbf a^T\HH\mathbf a+o(|\varepsilon|)$. Taking $\mathbf a$ to be the $i$th coordinate vector gives the comparison for $\HH_{ii}$.

Let $\Delta\HH=\HH(L',Q)-\HH(L,Q)$. Its two-by-two principal minors give
\begin{equation}
 \Delta\HH_{ii}\geq0,\qquad
 |\Delta\HH_{ik}|^2\leq\Delta\HH_{ii}\Delta\HH_{kk}.
 \unreferencedlabel{eq:crossbound}
 \notag
\end{equation}
The second inequality constrains the change in the angular velocity induced on one horizon by the spin of another. If either of the two diagonal coefficients is unchanged, the corresponding off-diagonal coefficient must also be unchanged. The sign of an individual off-diagonal coefficient, or of its change, is not fixed.

Because of Eq.~\eqref{eq:Hpd}, the inverse matrix is well defined. At linear order set
\begin{equation}
 \II=\HH^{-1},\qquad
 \boldsymbol J=\II\boldsymbol\Omega+o(|\Omega|).
 \unreferencedlabel{eq:inertia}
 \notag
\end{equation}
Inversion reverses positive-definite matrix order, so
\begin{equation}
 \II(L',Q)\leq\II(L,Q).
 \label{eq:Iorder}
\end{equation}
For fixed $\boldsymbol\Omega$, Eq.~\eqref{eq:Iorder} compares the quadratic form $\boldsymbol\Omega^T\II\boldsymbol\Omega$, the leading term of $\sum_i\Omega_iJ_i$. Strict attraction at nonzero spin need not make $\Delta\HH$ positive definite, since $D(\varepsilon\mathbf a)$ may vanish to order higher than two in some direction $\mathbf a$.

\subsection{Symmetric and antisymmetric binary directions}

For two equal charges, assume that reflection exchanges the two horizons and maps the chosen family at $(J_1,J_2)$ to the same family at $(J_2,J_1)$. Then
\begin{equation}
 \HH(L,Q)=
 \begin{pmatrix}a(L)&b(L)\\ b(L)&a(L)\end{pmatrix}.
 \unreferencedlabel{eq:binarymatrix}
 \notag
\end{equation}
The eigenvectors are $(1,1)$ and $(1,-1)$. The symmetric (corotating) perturbation $J_1=J_2=j$ has $\Omega_1=\Omega_2=(a+b)j+o(j)$; the antisymmetric (counterrotating) perturbation $J_1=-J_2=j$ has $\Omega_1=-\Omega_2=(a-b)j+o(j)$. Both coefficients are positive and nondecreasing with $L$:
\begin{equation}
 0<a(L)\pm b(L)\leq a(L')\pm b(L'),\qquad L'\geq L.
 \unreferencedlabel{eq:binarymodes}
 \notag
\end{equation}
In particular, $a(L')-a(L)\geq|b(L')-b(L)|$. Thus the change of the off-diagonal coefficient is bounded in absolute value by the change of the diagonal coefficient. These are statements at fixed individual charges on the extremal branch, so their comparison conditions differ from the fixed-mass double-Kerr examples of Ref.~\cite{CHR}.

\subsection{Quadratic force generated by small spins}

If the mixed derivatives permit a $C^2$ Taylor expansion of $\partial_{L_j}\cE$ in $J$---for example, if $\cE$ is $C^3$ jointly---then Eq.~\eqref{eq:responseexpansion} gives
\begin{align}
 \alpha_j&=-2J^T(\partial_{L_j}\HH)J+o(|J|^2),\notag\\
 \mu_j&=\tfrac12J^T(\partial_{L_j}\HH)J+o(|J|^2).
 \label{eq:spinforce}
\end{align}
Thus the quadratic force induced by rotation is determined by the separation derivative of the spin Hessian. Equation~\eqref{eq:forcesign} implies $\partial_{L_j}\HH\geq0$ wherever this derivative exists. The finite inequality \eqref{eq:Horder} remains valid without this stronger regularity.

The distinction between $\cE$ and $m$ also appears at quadratic order. Under the same stronger mixed regularity, Eqs.~\eqref{eq:potentialdefinition} and \eqref{eq:spinforce} give
\begin{equation}
 m=\sum_i Q_i+\frac12J^T\!\left(\HH-\sum_jL_j\partial_{L_j}\HH\right)\!J
       +o(|J|^2).
 \unreferencedlabel{eq:admhessian}
 \notag
\end{equation}
Consequently, positivity and comparison of $\HH$ are assertions about the corrected mass function, not about the angular momentum Hessian of the ADM mass.

For a single isolated extremal Kerr--Newman black hole,
\begin{equation}
 \begin{aligned}
 m_{\rm ext}(J,Q)
 &=\left[\frac{Q^2+\sqrt{Q^4+4J^2}}2\right]^{1/2}\\
 &=Q+\frac{J^2}{2Q^3}+O(J^4),
 \end{aligned}
 \unreferencedlabel{eq:KNcheck}
 \notag
\end{equation}
so $\HH=Q^{-3}$ and $\II=Q^3$ for $Q>0$. Comparison with these isolated values would require control of $\HH$ as the gaps tend to infinity, which is not assumed here.

\section{Discussion}

The first law identifies the differential of the ADM mass on a fixed-type parameter family. In the all-degenerate case, the renormalized energy proof separates variations of the axis constants from motion of the punctures. The latter produce a stress tensor boundary integral, and the correction between $\mathcal{E}/(8\pi)$ and $m$ converts the resulting position derivatives into the conical term. Section~\ref{sec:harmonicnondeg} proves the static reference energy variation by evaluating the moving pole terms and combining them with the horizon and reference variations. This proof assumes the differentiated pole and boundary estimates in Sec.~\ref{sec:input}; the remaining charged and parameter regularity estimates have not been proved here. Section~\ref{sec:iyerwald} outlines the Iyer--Wald approach under the additional boundary conditions in Appendix~\ref{app:covariantconditions}, for completeness of the literature review rather than as part of the proof. The static binary calculation checks the conical sign and normalization independently of uniform scaling.

The corrected mass function makes the second result a Hessian comparison. Theorem~1.1 of Ref.~\cite{HKWXev} gives $\partial_{L_j}\cE\geq0$, and $\cE(L,0,Q)=\sum_iQ_i$ supplies a common reference value. For configurations joined by a path satisfying Theorem~3, the difference of their corrected masses therefore has a minimum at $J=0$. Its Hessian is the matrix difference in Theorem~3. Square completion and the differentiated axis traces give strict positivity of each spin Hessian, while its principal minors and inverse yield the stated matrix corollaries.

The scope is fixed by the hypotheses in Sec.~\ref{sec:input}. Existence and the proved vacuum asymptotics and puncture position dependence do not replace the assumed differentiability in angular momenta and nondegenerate rod endpoints, or the remaining charged boundary estimates. Theorem~1.1 of Ref.~\cite{HKWXev} has its own asymptotic requirements throughout the comparison neighborhood. Separations are normalized Weyl gaps; the inequalities concern stationary angular momentum derivatives of $\cE$, not proper distances, ADM convexity, or dynamical stability. Collisions, changes of horizon type, and magnetic or charged axis sources require additional analysis.

\begin{acknowledgments}
Q. Han acknowledges the support of NSF Grant DMS-2305038. M. Khuri acknowledges the support of NSF Grant DMS-2405045. G. Weinstein acknowledges the support of ISF Grant 1119/26. J. Xiong acknowledges partial support from NSFC Grant 12325104.

OpenAI ChatGPT/Codex assisted with manuscript organization, literature comparison, and the development and checking of the displayed charged variational and response calculations, starting from the authors' research notes.
\end{acknowledgments}
\emph{Data availability.} No datasets were generated or analyzed in this theoretical study.

\appendix
\section{Boundary estimates and order of limits}
\label{app:boundary}

The integral conditions below supplement (B2). They are uniform on compact subsets of $\mathcal P$ and for parameter variations in bounded sets. Near a moving endpoint, $\delta$ denotes differentiation at fixed spatial coordinates $x$, while $D_s$ denotes differentiation at fixed centered coordinates $y=x-p(s)e_z$.

\subsection{Integral conditions for the linear estimates}
\label{app:linearconditions}

For $\delta=\partial_s$ at fixed spatial coordinates, define
\begin{align}
 \mathfrak j_n={}&e^{4u}\Theta_n(\delta v-\psi\delta\chi+\chi\delta\psi)\notag\\
 &+e^{2u}(\delta\chi\,\partial_n\chi+\delta\psi\,\partial_n\psi).
 \label{eq:potentialvariationcurrent}
\end{align}
At a nondegenerate pole, require
\begin{equation}
 \int_{S_{r_p}}|\mathfrak j_n|\,\dd S\leq Cr_p^\eta,
 \qquad \eta>0,
 \label{eq:polecurrentbound}
\end{equation}
uniformly up to the axis and for parameter variations in bounded sets. This condition on the potential variations is separate from the bound on $P$ and its parameter derivative.

We impose the following integral conditions in addition to (B1)--(B2). Set
\begin{align*}
 S_\rho&=\rho\bigl(\delta U\,\partial_\rho\overline U
              -\overline U\,\partial_\rho\delta U_*\bigr),\\
 K_\rho&=S_\rho+\rho\mathfrak j_\rho.
\end{align*}
For each compact interval $I\Subset\Gamma_j$ in an axis rod, define
\begin{align*}
 R_\rho={}&\rho e^{4u}\Theta_\rho
  (\delta v-\delta c_j-\psi\delta\chi+\chi\delta\psi)\\
 &+\rho e^{2u}\chi_\rho(\delta\chi-\delta d_j)
   +\rho e^{2u}\psi_\rho\delta\psi.
\end{align*}
We require
\begin{align*}
 \lim_{\rho\downarrow0}\int_I(|R_\rho|+|S_\rho|)\,\dd z&=0,\\
 \lim_{\rho\downarrow0}\int_I w\psi_z\,\dd z&=0.
\end{align*}
The functions $w(\rho,\cdot)$ and $A_t(\rho,\cdot)$ converge locally
uniformly to their axis traces as $\rho\downarrow0$. These traces extend
continuously to the finite rod endpoints with their prescribed values
and tend to zero at infinity; these limits and extensions are uniform
on compact parameter sets. The reconstruction identities then give
\begin{equation*}
 \lim_{\rho\downarrow0}\int_I K_\rho\,\dd z
 =-\frac12\delta c_j[w]_I-\delta d_j[A_t]_I,
\end{equation*}
where $[f]_I=f(b)-f(a)$ for $I=[a,b]$. On each compact interval $I$ in
the interior of a nondegenerate horizon rod, we instead require
\begin{equation*}
 \lim_{\rho\downarrow0}\int_I|K_\rho|\,\dd z=0.
\end{equation*}

For an endpoint or puncture $q$ and either adjoining rod $J$, put
\begin{equation*}
 J(q;a,t)=\{z\in J:a<|z-q|<t\}.
\end{equation*}
We require the endpoint and infinity conditions
\begin{equation*}
 \lim_{t\downarrow0}\sup_{0<a<t}\limsup_{\rho\downarrow0}
 \left|\int_{J(q;a,t)}K_\rho\,\dd z\right|=0,
\end{equation*}
\begin{equation*}
 \lim_{T\to\infty}\sup_{R>T}\limsup_{\rho\downarrow0}
 \left|\int_{J\cap\{T<|z|<R\}}K_\rho\,\dd z\right|=0,
\end{equation*}
with the second condition imposed on each unbounded axis rod. All limits
are uniform on compact parameter sets and for parameter variations in
bounded sets. The limits on compact intervals are also locally uniform
in the interval endpoints, while those endpoints remain in the interior
of the same rod. The cylinder radius tends to zero first, with the radii
$h$, $\varepsilon$, and $R$ fixed; we then take $h\to0$,
$\varepsilon\to0$, and $R\to\infty$, in that order. These endpoint
conditions concern only the cylinder integrals. The reference sphere
fluxes at nondegenerate poles and the translation fluxes at degenerate
punctures are retained.

At a degenerate puncture $p_i(s)$, write $D_s=\partial_s+\dot p_i\cdot\nabla$ for differentiation at fixed centered coordinates, as in Eq.~\eqref{eq:centeredvariation}. We assume that $D_sU$ is bounded and that the integrals of
\begin{align}
 r_i^2 e^{4u}\Theta_{r_i}
           (D_sv-\psi D_s\chi+\chi D_s\psi),\notag\\
 r_i^2 e^{2u}\left(\chi_{r_i}D_s\chi+\psi_{r_i}D_s\psi\right)
 \label{eq:puncturefluxrequirements}
\end{align}
against $\sin\theta_i\,\dd\theta_i$ tend to zero, including the polar limits. The Weyl profiles converge to their axis traces so that Eq.~\eqref{eq:stressflux} applies. These requirements concern the centered remainder; the nonzero translation flux is retained. At infinity we assume
\begin{equation*}
 \lim_{R\to\infty}\int_{S_R}\mathfrak j_n\,\dd S=0,
\end{equation*}
uniformly on compact parameter sets and for parameter variations in
bounded sets, with $n$ outward. The scalar outer-sphere currents tend
to zero by (B2) and the explicit reference potential.

For $\mathcal E$ on the all-degenerate family and $\widehat{\mathcal E}$
in general, impose the following criterion at each successive cutoff
limit. After taking the earlier limits, keep the remaining cutoffs fixed
and denote the energy by $F_\varepsilon$, with $\varepsilon$ the current
cutoff parameter ($R^{-1}$ at infinity). On a local parameter neighborhood
where the cutoffs are admissible, assume $F_\varepsilon\in C^1$, with
derivatives including the motion of the excised balls, and
\begin{equation*}
 F_\varepsilon\longrightarrow F,\qquad
 \dd F_\varepsilon\longrightarrow\omega,
\end{equation*}
pointwise and locally uniformly, respectively,
where $\omega$ is a continuous one-form. For a sufficiently small
parameter increment $\mathbf q$, the line-segment identity gives
\begin{align*}
 &F_\varepsilon(\mathbf p+\mathbf q)-F_\varepsilon(\mathbf p)\\
 &\quad=\int_0^1\dd F_\varepsilon(\mathbf p+t\mathbf q)[\mathbf q]\,\dd t\\
 &\quad\longrightarrow\int_0^1\omega(\mathbf p+t\mathbf q)[\mathbf q]\,\dd t.
\end{align*}
Thus $F\in C^1$ and $\dd F=\omega$. The convergence assumptions do not
follow from the finiteness established for $\mathcal E$ in
Ref.~\cite{KW}. At a nondegenerate pole, (B1)--(B2) give the integrable
bounds $|D_s(|\nabla\overline U|^2)|\leq Cr_p^{2\gamma-2}$ and
$|D_sP|\leq Cr_p^{-1}$. These justify differentiation in translated pole
neighborhoods, not the remaining boundary limits. The rod integrals and their first variations converge
after identification with fixed intervals.

For fixed extremal punctures, $U=\log r_i+O(1)$, $\delta U=O(1)$, and $\partial_{r_i}U=O(r_i^{-1})$ make the scalar flux in Eq.~\eqref{eq:energyboundary} vanish as $O(r_i)$. The potential fluxes vanish by Eq.~\eqref{eq:puncturefluxrequirements} in (B2).

The axis and infinity limits follow from the integral hypotheses above, together with the differentiated scalar asymptotics in (B2). For a position change $p_i(s)$, use local coordinates $y=x-p_i(s)$. The derivative at fixed $y$ is $\partial_sX_s+\dot p_i\cdot\nabla X_s$. Its boundary flux vanishes by the centered estimates. Combining the translation part of the field variation with motion of the domain gives the stress tensor and the nonzero flux in Eq.~\eqref{eq:stressflux}. In vacuum this displacement flux encodes the tangent-map parameter, as explained after Eq.~\eqref{eq:fullenergyvariation}.

For the square completion \eqref{eq:square}, use the boundary flux conditions in Sec.~\ref{sec:squarecompletion}. For Eq.~\eqref{eq:Hpositive}, local parameter differentiability of the fields and their spatial gradients gives pointwise limits of the square residuals divided by the spin parameter. Fatou's lemma supplies the displayed lower bound. Its left-hand side uses $C^2$ spin dependence of the scalar $\cE$, and strict positivity uses the preserved differentiated axis traces. Weighted $L^2$ differentiability of the square residuals would strengthen the lower bound to equality, but is not needed for the theorem.

\subsection{Conditions for the Iyer--Wald outline}
\label{app:covariantconditions}

The horizon identification of Appendix~\ref{app:horizon} is regular away from the polar cutoffs. On each compact subinterval of an axis rod the metric has the differentiated conical expansion used in Sec.~\ref{sec:covariant}, and the Maxwell potential satisfies Eq.~\eqref{eq:axisA} with differentiated coefficient bounds. The absolute integral of the charge two-form on each connecting cap tends to zero, uniformly on compact parameter sets and for parameter variations in bounded sets. In these cap estimates, $r$ denotes a physical radius centered at the corner and measured in the local regular conical spatial geometry, not an auxiliary Weyl cutoff radius. A sufficient estimate is a charge density bounded by $Cr^{-1}$ on caps of area at most $Cr^2$, measured in the local regular conical frame. Horizon integrals converge to the variations of the full cross-section quantities; degenerate horizon sections are retained rather than discarded as caps.

Let $r_\perp$ be proper radial distance to the axis. The local Maxwell expansion assumed here is
\begin{equation}
 \begin{split}
 \cA={}&a_t(r_\perp^2,z)\dd t+a_z(r_\perp^2,z)\dd z\\
       &+r_\perp a_{r_\perp}(r_\perp^2,z)\dd r_\perp+r_\perp^2a_\phi(r_\perp^2,z)\dd\phi,
 \end{split}
 \label{eq:axisA}
\end{equation}
with bounded coefficient functions and differentiated coefficients.

For the electromagnetic boundary estimates we use
\begin{equation}
 \boldsymbol k^{\rm EM}_\xi=-\frac1{4\pi}
 \left[(\iota_\xi\cA)\delta(\star F)
          +a\wedge\iota_\xi\star F\right].
 \label{eq:EMcharge}
\end{equation}
Here $a=\delta\cA$, and the variation of $\star F$ includes the variation of the metric.

The Maxwell expansion \eqref{eq:axisA} gives $(\star F)_{z\phi}$ and $\delta(\star F)_{z\phi}$ are $O(r_\perp^2)$, while $(\tau\cdot\star F)_z=O(1)$ and $(\tau\cdot\star F)_\phi=O(r_\perp^2)$. Substitution in Eq.~\eqref{eq:EMcharge} gives a vanishing Maxwell contribution on a shrinking axis tube.

At nondegenerate horizon-axis corners, the covariant boundary conditions require vanishing of the connecting-cap charge. With $r$ the same physical radius defined above, a sufficient version assumes $|\boldsymbol k_\tau|\leq Cr^{-1}$ in the regular conical frame on a cap with area at most $Cr^2$, uniformly on compact parameter sets and for parameter variations in bounded sets. The absolute integral is then $O(r)$. This bound is assumed for the differentiated metric and Maxwell fields at the corners. The axis tube is first shrunk on segments with endpoint cutoffs fixed, and those cutoffs are then removed. Degenerate throat cross-sections have finite limiting area; their horizon charges in Eq.~\eqref{eq:fullhorizon} must be retained instead of treating them as vanishing corner caps.

\section{The gravitational charge at a future horizon}
\label{app:horizon}

For reference, the boundary forms in the conventions used here are
\begin{align}
 \boldsymbol Q_\xi&=-\frac1{16\pi}\star\dd\xi^\flat
                  -\frac1{4\pi}(\iota_\xi\cA)\star F,\notag\\
 \boldsymbol\vartheta^{\rm EM}&=-\frac1{4\pi}a\wedge\star F,\notag\\
 \vartheta^{\rm GR}_{abc}&=\epsilon_{dabc}\theta^d,
 \qquad\theta^d=\frac{\nabla_eh^{de}-\nabla^dh}{16\pi}.
 \label{eq:noetherforms}
\end{align}
Here $h_{ab}=\delta g_{ab}$, $a=\delta\cA$, $h=g^{ab}h_{ab}$, and $h^{ab}=g^{ac}g^{bd}h_{cd}$.

For one horizon, suppress the component index and use Gaussian null coordinates with fixed background generator $\xi=\partial_V$:
\begin{equation}
 \begin{split}
 \dd s^2={}&2\dd V\dd r-2\kappa r\dd V^2
          +2r\beta_A\dd V\dd x^A\\
          &+\gamma_{AB}(r,x)\dd x^A\dd x^B+O(r^2).
 \end{split}
 \unreferencedlabel{eq:gnc}
 \notag
\end{equation}
The displayed remainders refer to the $VV$ and $VA$ coefficients. Identify the varied horizon with $r=0$ and choose $h_{Vr}=h_{rr}=h_{rA}=0$. Put $P_1=\partial_rh_{VV}|_0$ and $s_A=h_{VA}|_0$. Nullness gives $h_{VV}|_0=0$, and orthogonality of the varied generator $\xi+\delta\Omega\eta$ gives $s_A=-\delta\Omega\eta_A$. In a locally stationary identification gauge, Eq.~\eqref{eq:noetherforms} yields
\begin{align}
 16\pi\theta^r&=P_1-\beta_As^A+D_As^A,\notag\\
 \delta\boldsymbol Q^{\rm GR}_\xi\big|_H
 &=\frac{\kappa}{8\pi}\delta(\dd A)
                         -\frac{P_1-\beta_As^A}{16\pi}\dd A,\notag\\
 \boldsymbol k^{\rm GR}_\xi\big|_H
 &=\frac{\kappa}{8\pi}\delta(\dd A)
                              +\frac{D_As^A}{16\pi}\dd A.
 \unreferencedlabel{eq:gnccalculation}\notag
\end{align}
Since $\eta$ is an axial Killing field of the cross-section and $\delta\Omega$ is constant, $D_As^A=0$. For a nondegenerate horizon, the integrated gravitational contribution is $\kappa\delta A/(8\pi)$. For a degenerate horizon, the displayed local gravitational density vanishes and its integral is zero. Equation~\eqref{eq:EMcharge} gives the electromagnetic contribution $\Phi\delta Q$. The local horizon identification can be extended to the identity at infinity; the variation need not remain stationary in the interpolation region for the exact-background identity $\dd\boldsymbol k_\tau=0$ to hold.

\section{Check on the Bach--Weyl two body solution}
\label{sec:binarycheck}

Consider the Bach--Weyl static two-body solution \cite{BachWeyl1922}; an English translation of the original paper is available in Ref.~\cite{BachWeylEnglish}. We use the parametrization and first-law formulas of Ref.~\cite{KZ}. Let $M_1,M_2>0$ be the mass parameters. The two horizon rods have Weyl coordinate lengths $2M_1$ and $2M_2$. If their midpoint coordinates are $z_1<z_2$, define $d=z_2-z_1>M_1+M_2$. Thus $d$ is a coordinate separation, not a proper distance. Put $S=M_1+M_2$, $D=M_1-M_2$. The ADM mass is $m=S$ and the gap is $L=d-S$. The horizon data are
\begin{align}
 A_1&=16\pi M_1^2\frac{d+S}{d+D},&
 \kappa_1&=\frac{d+D}{4M_1(d+S)},\notag\\
 A_2&=16\pi M_2^2\frac{d+S}{d-D},&
 \kappa_2&=\frac{d-D}{4M_2(d+S)}.
 \unreferencedlabel{eq:staticbinarydata}\notag
\end{align}
The conical quantities in our convention are
\begin{equation}
 \mu=\frac{M_1M_2}{d^2-S^2},\qquad
 \lambda=(d-S)\frac{d^2-S^2}{d^2-D^2}.
 \unreferencedlabel{eq:staticbinarycone}
 \notag
\end{equation}
Differentiating these expressions in the independent variables $(M_1,M_2,d)$ gives
\begin{equation}
 \frac{\kappa_1}{8\pi}\dd A_1+\frac{\kappa_2}{8\pi}\dd A_2
          -\lambda\dd\mu=\dd M_1+\dd M_2.
 \unreferencedlabel{eq:staticbinarycheck}
 \notag
\end{equation}
In particular, at fixed masses the coefficient of $\dd d$ from the two area terms is
\begin{equation}
 -\frac{2M_1M_2d}{(d+S)(d^2-D^2)}
       =\lambda\,\partial_d\mu.
 \unreferencedlabel{eq:staticbinaryR}
 \notag
\end{equation}
This verifies the sign of the conical term independently of the scaling variation.

\end{document}